\documentclass[useAMS,usenatbib]{mn2e}
\pdfoutput=1

\usepackage{times}
\usepackage{graphicx}
\usepackage{xcolor}
\usepackage{amssymb}
\usepackage{amsmath}
\usepackage{siunitx}
\usepackage{multirow}
\usepackage{booktabs}
\usepackage{csvsimple}

\title[Balancing the Energy Budget]{Balancing the Energy Budget of Short-Period Giant Planets: Evidence for Reflective Clouds and Optical Absorbers}
\author[J. C. Schwartz and N. B. Cowan]{J. C. Schwartz$^{1}$\thanks{E-mail: joelschwartz2011@u.northwestern.edu (JCS)} and N. B. Cowan$^{1, 2}$\thanks{E-mail: ncowan@amherst.edu (NBC)}\\
$^{1}$Department of Physics \& Astronomy, Northwestern University, 2145 Sheridan Road, Evanston, IL, 60208, USA\\
$^{2}$Department of Physics \& Astronomy, Amherst College, Amherst, MA, 01002, USA}
\begin{document}

\date{Submitted to MNRAS}

\pagerange{\pageref{firstpage}--\pageref{lastpage}} \pubyear{2015}

\maketitle

\label{firstpage}

\begin{abstract}
We consider fifty transiting short-period giant planets for which eclipse depths have been measured at multiple infrared wavelengths. The aggregate dayside emission spectrum of these planets exhibits no molecular features, nor is brightness temperature greater in the near-infrared. We combine brightness temperatures at various infrared wavelengths to estimate the dayside effective temperature of each planet. We find that dayside temperatures are proportional to irradiation temperatures, indicating modest Bond albedo and no internal energy sources, plus weak evidence that dayside temperatures of the hottest planets are disproportionately high. We place joint constraints on Bond albedo, $A_{B}$, and day-to-night transport efficiency, $\varepsilon$, for six planets by combining thermal eclipse and phase variation measurements (HD~149026b, HD~189733b, HD~209458b, WASP-12b, WASP-18b, and WASP-43b). We confirm that planets with high irradiation temperatures have low heat transport efficiency, and that WASP-43b has inexplicably poor transport; these results are statistically significant even if the precision of single-eclipse measurements has been overstated by a factor of three. Lastly, we attempt to break the $A_{B}$--$\varepsilon$ degeneracy for nine planets with both thermal and optical eclipse observations, but no thermal phase measurements. We find a systematic offset between Bond albedos inferred from thermal phase variations ($A_{B} \approx 0.35$) and geometric albedos extracted from visible light measurements ($A_{g} \approx 0.1$). These observations can be reconciled if most hot Jupiters have clouds that reflect 30--50 per cent of incident near-infrared radiation, as well as optical absorbers in the cloud particles or above the cloud deck.
\end{abstract}


\section{Introduction}
\label{sec:intro}
Mature planets on short-period orbits have energy budgets dominated by incoming radiation rather than internal heat. Their atmospheric temperatures are therefore a function of both the absorption of incident stellar energy and its transport before re-emission into space.

Absorbed energy is solely a matter of incident stellar flux and the planet's Bond albedo, $A_{B}$. Ironically, it is difficult to constrain Bond albedo (the fraction of stellar energy that is reflected) through observations of reflected light. In order to convert an optical geometric albedo (visible light reflected towards the illuminating star) into a Bond albedo, one must make assumptions about a planet's reflectance spectrum, scattering phase function, and spatial inhomogeneity \citep{hanel2003exploration}. Bond albedo is more readily obtained from thermal measurements via energy balance.

Heat transport is more complicated, but tends to move energy from hot to cold: vertically upwards, from equator to pole (for planets with small obliquity), from summer hemisphere to winter hemisphere (for planets with non-zero obliquity), and from day to night (for planets with slow rotation). Due to strong tides, short-period planets are expected to have zero obliquity and slow rotation, and most are on circular orbits. Hot Jupiters on circular orbits are further expected to be tidally-locked, with one side permanently facing the host star and the other forever dark. As such, the atmosphere tends to transport heat from the dayside to the nightside, and from equatorial regions to the poles \citep[for a recent review of hot Jupiter atmospheric dynamics, see][]{heng2014atmospheric}.

The atmospheric temperature of a planet is generally a function of four variables: longitude, latitude, pressure (or height), and time. The time-dependence can usually be neglected for hot Jupiters because they exhibit minimal weather \citep{Agol_2010, Knutson_2011, Wong_2014}. Moreover, hot Jupiters on circular orbits are expected to have a 3D fixed temperature structure with respect to the sub-stellar location, regardless of whether they are tidally locked \citep{Rauscher_2014, Showman_2015}. This motivates using a star-based coordinate system with the prime meridian facing the star, and allows us to use orbital phase as a proxy for longitude \citep{Cowan_2008}. The latitudinal temperature-dependence of a hot Jupiter is inaccessible unless one can measure higher-order phase modulation \citep{Cowan_2013} or utilize occultation mapping \citep{Majeau_2012, deWit_2012}. Finally, the vertical temperature structure is in principle accessible via emission spectroscopy: wavelengths at which the atmosphere is relatively transparent will probe deeper layers, and vice versa.

Multi-wavelength thermal phase variations of a hot Jupiter on a circular orbit therefore amount to brightness temperature measurements as a function of orbital phase and wavelength \citep[e.g.,][]{Stevenson_2014}. If one is solely interested in the global properties of the planet---namely Bond albedo and day-to-night heat transport---then one can further simplify the problem by combining brightness temperatures at each orbital phase to obtain a bolometric flux, and hence an effective temperature \emph{at that phase}. Note that Solar System planets tend to have effective temperatures that are either uniform from any vantage point, or which vary based on the latitude of the observer due to imperfect poleward heat transport. For short-period exoplanets, on the other hand, the principal temperature gradient is between day and night, and dayside effective temperatures are often hundreds to thousands of degrees greater than their nightside counterparts. The final simplification we make is therefore to treat the planet as two horizontally isothermal hemispheres: a dayside and a nightside. The effective temperatures of each hemisphere are simply the weighted mean of the atmospheric temperatures on that side of the planet.

\subsection{Previous Work}
\citet{cowan2011statistics} used broadband infrared eclipse measurements of 24 hot Jupiters to demonstrate that they have generally low Bond albedos ($A_B < 0.5$), and that the hottest planets have extremely low albedos and/or poor day--night heat transport efficiency, $\varepsilon$.

It is possible to break the albedo-transport degeneracy by combining dayside thermal constraints with measurements of either nightside thermal emission or dayside reflected light. \citet{cowan2007hot} used an $8.0~\mu$m eclipse depth and phase amplitude from the Spitzer Space Telescope, combined with an optical eclipse measurement from the MOST satellite, to constrain the energy budget of HD~209458b; they inferred a small Bond albedo (absorption of almost all light that shines on it) and a high day-night heat transport efficiency (nightside not much cooler than the dayside).

\citet{cowan2012thermal} used 8 infrared dayside and 2 mid-infrared nightside measurements to constrain the albedo and recirculation of WASP-12b; they found the planet has a modest Bond albedo ($\sim 0.25$) and low heat transport efficiency ($\lesssim 0.1$).

\citet{stevenson2014thermal} used phase-resolved emission spectroscopy taken with WFC3 from the Hubble Space Telescope to map the atmospheric thermal structure of WASP-43b, finding low Bond albedo ($0.06$--$0.25$) and no heat redistribution \citep[for recent reviews of exoplanet atmospheric observations, please see][]{burrows2014spectra, bailey2014dawes}.

Our work is organized as follows: in Section~\ref{sec:dayEB} we use published eclipse depths at multiple infrared wavelengths to infer effective dayside temperatures for fifty planets, with more than twice the data as \cite{cowan2011statistics}. In Section~\ref{sec:globalEB} we consider the subset of planets for which we can break the albedo-recirculation degeneracy. We first tackle the six planets with thermal measurements of both eclipses and phase variations (Section~\ref{sec:daynight}), then the nine planets for which reflected light measurements are available in addition to dayside thermal constraints (Section~\ref{sec:dayalb}). We discuss our results in Section~\ref{sec:discuss}.

\section{Dayside Energy Budget}
\label{sec:dayEB}
Inferring the effective dayside temperature of a planet requires combining eclipse depths at thermal wavelengths, which we define as those longward of $0.8~\mu $m. We have updated the sample from \cite{hansen2014broadband}, now including fifty planets with a minimum of two published infrared eclipse measurements \citep[additions include HAT-P-19b, HAT-P-20b, HAT-P-32b, WASP-6b, WASP-26b, WASP-39b;][Kammer et al. in prep.]{mahtani2013warm,deming2014spitzer,zhao2014characterization}. Our data predominantly consist of broadband photometry, but we also include spectroscopic emission measurements when they are at complementary wavelengths \citep[e.g.][]{ranjan2014atmospheric, wilkins2014emergent, crouzet2014water, stevenson2014thermal}. The planets from our sample are listed in Table \ref{tab:allplanetsamp}.

\begin{table}
	\centering
	\caption{Short-period giant planets with a minimum of two published eclipse observations at infrared wavelengths (non-detections are not included.)}
	\label{tab:allplanetsamp}
	\begin{tabular}{c c}
		\toprule
		\multicolumn{2}{c}{\bfseries Infrared Multi-Eclipse Planets}\\
		\midrule
		\begin{filecontents*}{table1.csv}
planeta,planetb
CoRoT-1b,TrES-2b
CoRoT-2b,TrES-3b
GJ 436b,TrES-4b
HAT-P-1b,WASP-1b
HAT-P-2b,WASP-2b
HAT-P-3b,WASP-3b
HAT-P-4b,WASP-4b
HAT-P-6b,WASP-5b
HAT-P-7b,WASP-6b
HAT-P-8b,WASP-8b
HAT-P-12b,WASP-12b
HAT-P-19b,WASP-14b
HAT-P-20b,WASP-17b
HAT-P-23b,WASP-18b
HAT-P-32b,WASP-19b
HD 149026b,WASP-24b
HD 189733b,WASP-26b
HD 209458b,WASP-33b
KELT-1b,WASP-39b
Kepler-5b,WASP-43b
Kepler-6b,WASP-48b
Kepler-12b,XO-1b
Kepler-13Ab,XO-2b
Kepler-17b,XO-3b
TrES-1b,XO-4b
\end{filecontents*}
\csvreader[head to column names, late after line=\\]{table1.csv}{}{\planeta & \planetb}
		\bottomrule
	\end{tabular}
\end{table}

\subsection{Brightness Temperatures}
\label{sec:Tb}
Thermal emission at different wavelengths originates from different layers in the planet's atmosphere, which have different temperatures. One can define a brightness temperature at each observed wavelength, $T_{b} (\lambda)$: this is the temperature that a blackbody must have in order to emit at the same intensity as the planet.

Inverting the Planck function, we obtain the following expression for brightness temperature \citep{cowan2011statistics}:
\begin{equation}
\label{eq:bright}
T_{b} (\lambda) = \frac{hc}{\lambda k} \left[ \textrm{log}\left(1 + \frac{e^{hc/\lambda k T_{*}} - 1}{\psi (\lambda)}\right) \right]^{-1},
\end{equation}
where $\psi(\lambda)$ is the relative intensity of the planet to that of its host star and $T_{*}$ is the stellar effective temperature, meaning we treat the star as a blackbody. For dayside measurements, $\psi(\lambda)$ is the ratio of eclipse depth to transit depth, $\delta_\text{ecl} / \delta_\text{tr}$, while for nightside measurements it is the ratio of nigthside flux to transit depth, $(\delta_\text{ecl} - \delta_\text{var}) / \delta_\text{tr}$, where $\delta_\text{var}$ is the phase variation amplitude. Published data therefore allow us to compute dayside (and, when appropriate, nightside) brightness temperatures for each waveband in which a planet has been observed.

\subsection{Aggregate Emission Spectrum}
\label{sec:aggspec}
The broadband emission spectra of most individual planets are consistent with isothermal atmospheres \citep{hansen2014broadband}. It is possible, however, to construct an aggregate emission spectrum of all fifty planets in the hopes of revealing molecular absorption features too faint to detect in any individual planet's spectrum.

If some planets have temperature inversions while others do not, this type of averaging could actually wash out molecular signatures, which would appear in absorption for some planets and emission for others. However, the first and most statistically significant case of a hot Jupiter temperature inversion \citep{knutson20083} has not been borne out by new measurements nor reanalysis of the originals \citep{zellem20144,diamondlowe2014new, Schwarz_2015}. Moreover, a systematic study of \emph{Spitzer} eclipse measurements found that they have not been as accurate as advertised \citep{hansen2014broadband}, suggesting the temperature inversions reported in most hot Jupiter atmospheres may simply be due to confirmation bias.

\begin{figure}
	\centering
	\includegraphics[width=1.0\linewidth]{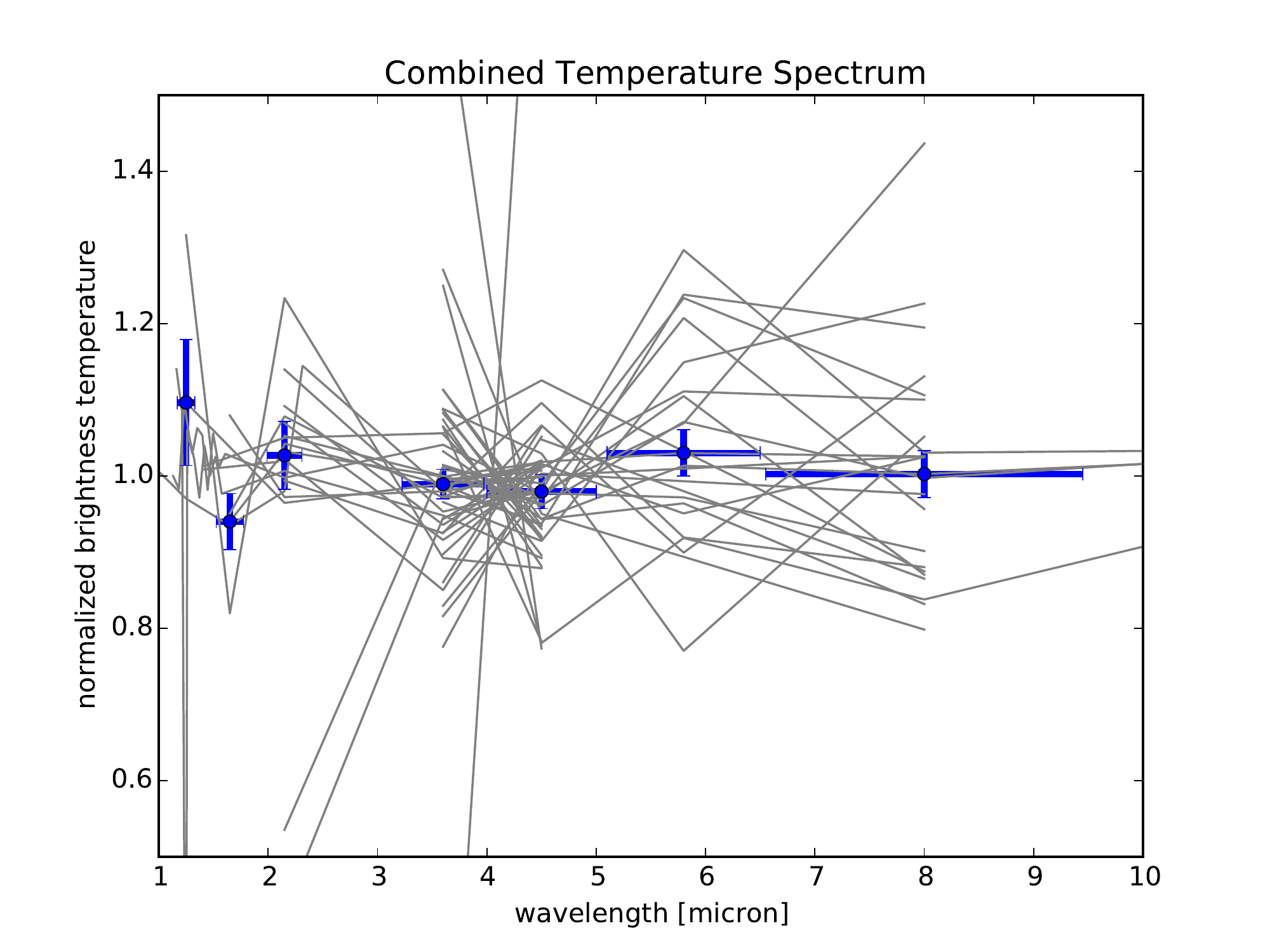}
	\caption{Average broadband emission spectrum for fifty short-period giant planets (blue), plotted with the emission spectra of individual planets (gray). The spectrum of each planet is normalized to its mean brightness temperature. The aggregate spectrum is the median normalized brightness temperature at each wavelength, where the uncertainty bars denote the uncertainty on the mean (as opposed to the standard variation of the spectra at that wavelength.)}
	\label{fig:Tb_spectrum}
\end{figure}

The aggregate emission spectrum for the fifty hot Jupiters is shown in Figure~\ref{fig:Tb_spectrum}. We normalize the brightness temperature spectrum of each planet in our sample, then determine the median and uncertainty on the mean at each wavelength. This ``stacking'' is only useful, however, at wavelengths for which there are observations of many planets (currently 1.15, 1.65, 2.25, 3.6, 4.5, 5.8, and 8.0 microns).

\begin{figure}
	\centering
	\includegraphics[width=1.0\linewidth]{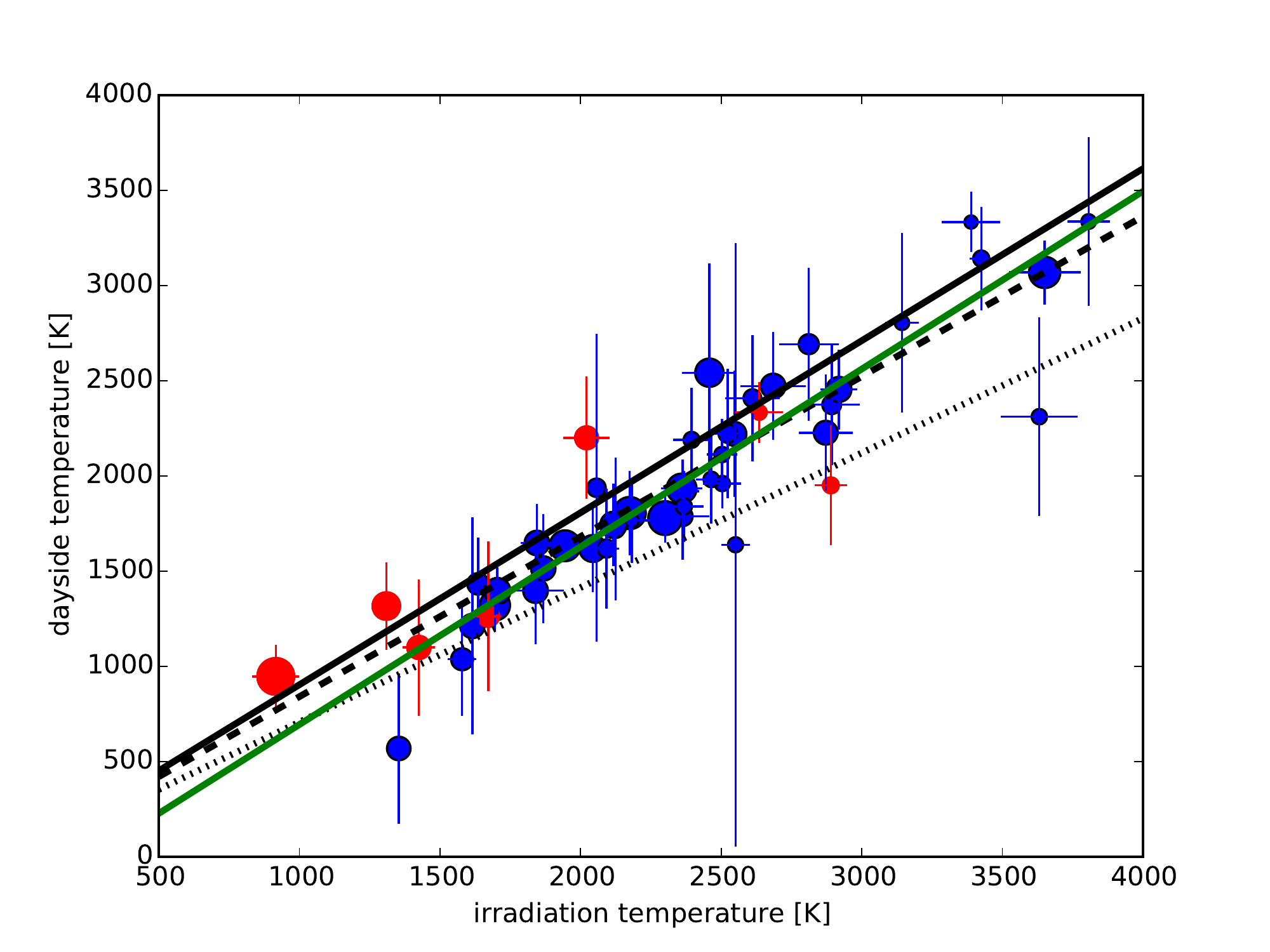}
	\caption{Dayside effective temperature versus irradiation temperature for all giant planets with multiple dayside infrared eclipses, estimated by Monte Carlo using hybrid EWM-PWM calculation and inflating observational uncertainties by $f_\text{sys} = 3$ where applicable \citep{hansen2014broadband}. Uncertainty bars for both temperatures are shown, while dot size is proportional to the fraction of expected planetary emission that falls within observed wavebands. Red symbols denote eccentric planets ($e > 0.1$). The solid, dashed, and dotted lines correspond respectively to maximum dayside temperature, $T_{d} = (2/3)^{1/4} T_{0}$, uniform dayside with zero nightside temperature, $T_{d} = (1/2)^{1/4} T_{0}$, and equilibrium temperature, $T_{d} = (1/4)^{1/4} T_{0}$. The trend line is shown in green; it suggests that hotter planets have disproportionately hot daysides.}
	\label{fig:TdToscatter}
\end{figure}

There are no significant features in the average spectrum, not even the trend towards higher brightness temperatures at shorter wavelengths reported by \cite{burrows2014spectra}. It is worth noting that Figure~4 of \citet{burrows2014spectra} used data from fewer planets, and was normalized differently: the \emph{equilibrium temperature} of each planet was divided out, rather than its actual \emph{dayside effective temperature}. Recall that a planet's equilibrium temperature is what one would expect for a planet with zero Bond albedo and uniform temperature; it is merely a convenient theoretical quantity proportional to the irradiation temperature, $T_0$, that we utilize in this work. The dayside effective temperature, on the other hand, is the weighted mean brightness temperature of the planet's dayside, as described in Section~\ref{sec:Teff}. Most hot Jupiters have dayside effective temperatures greater than their equilibrium temperature (dotted line in Figure~\ref{fig:TdToscatter}) due to imperfect day--night heat transport. As a result, the dayside of a hot Jupiter emits somewhat more in the mid-IR---and considerably more in the near-IR---than one would predict based on its equilibrium temperature. In any case, we agree with \cite{burrows2014spectra} that there are no signs of molecular absorption features in the aggregate spectrum. Since molecules are undoubtedly present in the atmospheres of exoplanets, we conclude that their absorption features are being muted by vertically isothermal atmospheres, optically thick cloud, or both.

\subsection{Effective Temperatures}
\label{sec:Teff}
While brightness temperatures of brown dwarfs are strongly wavelength-dependent in the near-infrared \citep{faherty2014signatures, biller2013weather}, the external, asymmetric heating experienced by hot Jupiters produces dayside atmospheres that are relatively isothermal in the vertical direction \citep{Fortney_2006}. This results in relatively featureless dayside emission spectra, which are amenable to model-independent estimates of bolometric flux and hence effective dayside temperature. If the nightsides of hot Jupiters have greater vertical temperature structure, then nightside effective temperature estimates are less reliable.

There is no universal way to derive the effective temperature of a planet from a collection of brightness temperatures. We therefore consider two methods with different physical motivations. In the first method, each brightness temperature is weighted by the inverse square of its respective uncertainty ($\omega_{i} = 1/\sigma_{i}^{2}$), so eclipse depths with small relative uncertainties contribute more to the inferred effective temperature. We call this the error-weighted mean (EWM) effective temperature. This method has the advantage of being robust to occasional outlier eclipse depths, but implicitly assumes that short-period planets have Planck-like broadband spectra.

The second method, which we only apply to dayside measurements, weighs the brightness temperatures by the expected integrated power in that waveband: $\omega_{i} = P_{i} = \int_{\lambda_{1}}^{\lambda_{2}} B(T_\text{est}, \lambda) d\lambda$, where $B(T_\text{est}, \lambda)$ is the Planck function for the planet's estimated dayside effective temperature. To bypass an iterative solution, we adopt $T_\text{est} = (\frac{1}{2})^{1/4} T_{0}$, where $T_{0}$ is the irradiation temperature: $T_{0} \equiv T_{*} \sqrt{R_{*}/a}$ where $R_{*}$ is the stellar radius and $a$ is the semi-major axis. This is an excellent match (the dashed line in Figure~\ref{fig:TdToscatter}) to the actual dayside effective temperature of most planets in our sample. We call this the power-weighted mean (PWM); it is identical in spirit to the linear interpolation method of \citet{cowan2011statistics}, but is easier to implement and runs faster. The PWM gives more weight to measurements near the peak of the planet's Planck function and should, in the limit of high spectral coverage, produce accurate effective temperatures even if planets have broadband spectral features.

Note that both EWM and PWM are biased in favor of broadband measurements: these observations tend to have smaller uncertainties, and they capture more of the planet's expected blackbody emission. The two methods produce generally consistent effective temperature estimates, which is a testament to the fact that most current dayside emission spectra are approximately Planck-like.

We use a $10^4$-step Monte Carlo analysis to estimate uncertainties in dayside effective temperatures. At each step in the Monte Carlo, we randomly vary the stellar effective temperature, transit depth, eclipse depth, and scaled semi-major axis, $a_{*} \equiv a/R_{*}$, according to their uncertainties. We use the published uncertainties for all of the above, except for single-epoch broadband eclipse depths where we inflate the published uncertainty by the factor $f_\text{sys} = 3$ \citep[such measurements have historically been less accurate than advertised;][]{hansen2014broadband}. We also compute each planet's irradiation and brightness temperatures (following Equation~\ref{eq:bright}). We then estimate each planet's dayside effective temperature; to hedge our bets, we use the EWM for half of the MC steps, and the PWM for the other half. The resulting relationship between $T_{0}$ and $T_{d}$ is shown in Figure~\ref{fig:TdToscatter}. The median property is plotted, and uncertainty bars denote standard deviation from the MC.

\section{Global Energy Budget}
\label{sec:globalEB}
Based on dayside effective temperatures alone, one cannot simultaneously specify Bond albedo and heat recirculation efficiency. This degeneracy can be broken by supplementing thermal eclipses with one of two measurement types: phase variations at infrared wavelengths, or eclipse depths at visible wavelengths. Table~\ref{tab:alt2edpvdata} lists the published data for the fifteen planets which fall into one or both of these categories: observed wavelength and bandwidth, eclipse depths, and phase amplitudes. Cyan-colored entries are measurements exempt from the $f_\text{sys} = 3$ uncertainty inflation of \citet{hansen2014broadband}.

\begin{table*}
	\begin{minipage}{165mm}
		\tiny
		\centering
		\caption{Eclipse depths and phase amplitudes for our restricted planetary samples where degeneracy between albedo and heat recirculation can be resolved. Observations are denoted by central wavelength and bandwidth; measurements in cyan are exempt from the uncertainty inflation $f_\text{sys} = 3$ of \protect\citet{hansen2014broadband}.}
		\label{tab:alt2edpvdata}
		\begin{tabular}{c S[scientific-notation=false, table-parse-only] S[table-format=1.4(3)e1] S[table-format=1.3(3)e1] c c S[scientific-notation=false, table-parse-only] S[table-format=1.3(3)e1] S[table-format=1.4(3)e1] @{}l}
			\toprule
			Planet & {Wavelength ($\mu$m)} & {Eclipse Depth} & {Phase Amplitude} & \textemdash\textemdash\textemdash & Planet & {Wavelength ($\mu$m)} & {Eclipse Depth} & {Phase Amplitude} & \\
			\midrule
			\begin{filecontents*}{table2.csv}
exo,wave,band,ed,edu,pa,pau,nsed,nspa,exob,waveb,bandb,edb,edub,pab,paub,nsedb,nspab
CoRoT-1b,0.6 \pm 0.42,0,0.00016 \pm 0.00006,0,,0,cyan,black,Kepler-7b,0.65 \pm 0.4,0,0.0000387 \pm 0.00000835,0,0.000048 \pm 0.000013,0,cyan,cyan
,1.65 \pm 0.25,0,0.00145 \pm 0.00049,0,,0,black,black,,3.6 \pm 0.75,0,0.000308 \pm 0.000103,0,,0,black,black
,2.1 \pm 0.02,0,0.0028 \pm 0.0005,0,,0,black,black,,4.5 \pm 1,0,0.000505 \pm 0.000168,0,,0,black,black
,2.15 \pm 0.32,0,0.00319 \pm 0.000405,0,,0,black,black,Kepler-13Ab,0.65 \pm 0.4,0,0.000172 \pm 0.0000018,0,0.000152 \pm 0.00000105,0,cyan,cyan
,3.6 \pm 0.75,0,0.00415 \pm 0.00042,0,,0,black,black,,2.15 \pm 0.32,0,0.00122 \pm 0.00051,0,,0,black,black
,4.5 \pm 1,0,0.00482 \pm 0.00042,0,,0,black,black,,3.6 \pm 0.75,0,0.00156 \pm 0.00031,0,,0,black,black
CoRoT-2b,0.6 \pm 0.42,0,0.00006 \pm 0.00002,0,,0,cyan,black,,4.5 \pm 1,0,0.00222 \pm 0.00023,0,,0,black,black
,1.4 \pm 0.6,0,0.000395 \pm 0.000057,0,,0,cyan,black,TrES-2b,0.65 \pm 0.4,0,0.0000077 \pm 0.0000018,0,0.0000041 \pm 0.00000105,0,cyan,cyan
,1.65 \pm 0.25,0,0.00085 \pm 0.000283,0,,0,black,black,,2.15 \pm 0.32,0,0.00062 \pm 0.00012,0,,0,black,black
,2.15 \pm 0.32,0,0.0016 \pm 0.0009,0,,0,black,black,,3.6 \pm 0.75,0,0.00127 \pm 0.00021,0,,0,black,black
,3.6 \pm 0.75,0,0.00355 \pm 0.0002,0,,0,black,black,,4.5 \pm 1,0,0.0023 \pm 0.00024,0,,0,black,black
,4.5 \pm 1,0,0.00475 \pm 0.00019,0,,0,black,black,,5.8 \pm 1.4,0,0.00199 \pm 0.00054,0,,0,black,black
,8 \pm 2.9,0,0.00409 \pm 0.0008,0,,0,black,black,,8 \pm 2.9,0,0.00359 \pm 0.0006,0,,0,black,black
HAT-P-7b,0.65 \pm 0.4,0,0.0000712 \pm 0.0000015,0,0.0000733 \pm 0.0000027,0,cyan,cyan,WASP-12b,0.9 \pm 0.15,0,0.00136 \pm 0.000136,0,,0,black,black
,3.6 \pm 0.75,0,0.00098 \pm 0.00017,0,,0,black,black,,1.04 \pm 0.12,0,0.00109 \pm 0.00014,0,,0,black,black
,4.5 \pm 1,0,0.00159 \pm 0.00022,0,,0,black,black,,1.25 \pm 0.16,0,0.00139 \pm 0.0003,0,,0,black,black
,5.8 \pm 1.4,0,0.00245 \pm 0.00031,0,,0,black,black,,1.38 \pm 0.55,0,0.00158 \pm 0.000039,0,,0,cyan,black
,8 \pm 2.9,0,0.00225 \pm 0.00052,0,,0,black,black,,1.4 \pm 0.6,0,0.00174 \pm 0.000017,0,,0,cyan,black
HD 149026b,3.6 \pm 0.75,0,0.0004 \pm 0.00003,0,,0,black,black,,1.65 \pm 0.25,0,0.00191 \pm 0.0002,0,,0,black,black
,4.5 \pm 1,0,0.00034 \pm 0.00006,0,0.000223 \pm 0.000058,0,cyan,cyan,,2.15 \pm 0.32,0,0.00296 \pm 0.00014,0,,0,cyan,black
,5.8 \pm 1.4,0,0.00044 \pm 0.0001,0,,0,black,black,,2.22 \pm 0.034,0,0.00301 \pm 0.00046,0,,0,black,black
,8 \pm 2.9,0,0.00037 \pm 0.00008,0,0.00023 \pm 0.00007,0,cyan,black,,2.32 \pm 0.027,0,0.0045 \pm 0.0006,0,,0,black,black
,16 \pm 5,0,0.00085 \pm 0.00032,0,,0,black,black,,3.6 \pm 0.75,0,0.00419 \pm 0.00044,0,0.0032 \pm 0.00033,0,cyan,cyan
HD 189733b,0.37 \pm 0.16,0,0.000126 \pm 0.000037,0,,0,cyan,black,,4.5 \pm 1,0,0.00429 \pm 0.00033,0,0.00392 \pm 0.00016,0,cyan,cyan
,0.51 \pm 0.12,0,0.000001 \pm 0.000034,0,,0,cyan,black,,5.8 \pm 1.4,0,0.00696 \pm 0.0006,0,,0,black,black
,1.4 \pm 0.6,0,0.000096 \pm 0.000039,0,,0,cyan,black,,8 \pm 2.9,0,0.00696 \pm 0.00096,0,,0,black,black
,2.15 \pm 0.32,0,0.0002 \pm 0.0002,0,,0,black,black,WASP-18b,3.6 \pm 0.75,0,0.00304 \pm 0.00026,0,0.00296 \pm 0.00011,0,cyan,black
,3.6 \pm 0.75,0,0.00147 \pm 0.00004,0,0.00124 \pm 0.000061,0,cyan,black,,4.5 \pm 1,0,0.00379 \pm 0.00021,0,0.00366 \pm 0.00009,0,cyan,black
,4.5 \pm 1,0,0.00179 \pm 0.000038,0,0.000982 \pm 0.000089,0,cyan,black,,5.8 \pm 1.4,0,0.0037 \pm 0.0003,0,,0,black,black
,5.8 \pm 1.4,0,0.0031 \pm 0.00034,0,,0,black,black,,8 \pm 2.9,0,0.0041 \pm 0.0002,0,,0,black,black
,6.45 \pm 2.1,0,0.0022 \pm 0.000062,0,,0,cyan,black,WASP-19b,0.685 \pm 0.53,0,0.00039 \pm 0.00019,0,,0,cyan,black
,8 \pm 2.9,0,0.00344 \pm 0.00036,0,,0,black,black,,0.91 \pm 0.2,0,0.0008 \pm 0.00029,0,,0,black,black
,10.5 \pm 6,0,0.00356 \pm 0.000067,0,,0,cyan,black,,1.29 \pm 0.08,0,0.00083 \pm 0.00039,0,,0,cyan,black
,16 \pm 5,0,0.00551 \pm 0.0003,0,,0,black,black,,1.6 \pm 0.4,0,0.00186 \pm 0.00014,0,,0,cyan,black
,24 \pm 9,0,0.00536 \pm 0.00027,0,0.0013 \pm 0.0003,0,cyan,black,,1.65 \pm 0.25,0,0.00276 \pm 0.00044,0,,0,black,black
HD 209458b,0.5 \pm 0.3,0,0.000007 \pm 0.000009,0,,0,cyan,black,,2.15 \pm 0.32,0,0.00287 \pm 0.0002,0,,0,black,black
,2.15 \pm 0.32,0,0.00015 \pm 0.00015,0,,0,black,black,,3.6 \pm 0.75,0,0.00483 \pm 0.00025,0,,0,black,black
,3.6 \pm 0.75,0,0.00094 \pm 0.00009,0,,0,black,black,,4.5 \pm 1,0,0.00572 \pm 0.0003,0,,0,black,black
,4.5 \pm 1,0,0.00139 \pm 0.0000705,0,0.00109 \pm 0.000115,0,cyan,black,,5.8 \pm 1.4,0,0.0065 \pm 0.0011,0,,0,black,black
,5.8 \pm 1.4,0,0.00301 \pm 0.00043,0,,0,black,black,,8 \pm 2.9,0,0.0073 \pm 0.0012,0,,0,black,black
,8 \pm 2.9,0,0.0024 \pm 0.00026,0,0.00075 \pm 0.000375,0,black,black,WASP-43b,1.4 \pm 0.6,0,0.000461 \pm 0.000005,0,0.000468 \pm 0.000004,0,cyan,cyan
,24 \pm 9,0,0.00338 \pm 0.00026,0,,0,cyan,black,,1.65 \pm 0.25,0,0.00103 \pm 0.00017,0,,0,black,black
Kepler-5b,0.65 \pm 0.4,0,0.0000186 \pm 0.0000036,0,0.0000193 \pm 0.0000058,0,cyan,cyan,,2.15 \pm 0.32,0,0.00181 \pm 0.00027,0,,0,black,black
,3.6 \pm 0.75,0,0.00103 \pm 0.00017,0,,0,black,black,,3.6 \pm 0.75,0,0.00347 \pm 0.00013,0,,0,black,black
,4.5 \pm 1,0,0.00107 \pm 0.00015,0,,0,black,black,,4.5 \pm 1,0,0.00382 \pm 0.00015,0,,0,black,black
Kepler-6b,0.65 \pm 0.4,0,0.0000111 \pm 0.000004,0,0.0000172 \pm 0.0000043,0,cyan,cyan,,,0,,0,,0,black,black
,3.6 \pm 0.75,0,0.00069 \pm 0.00027,0,,0,black,black,,,0,,0,,0,black,black
,4.5 \pm 1,0,0.00151 \pm 0.00019,0,,0,black,black,,,0,,0,,0,black,black
\end{filecontents*}
\csvreader[head to column names, late after line=\\]{table2.csv}{}{\exo & \wave & \color{\nsed} \ed & \color{\nspa} \pa & & \exob & \waveb & \color{\nsedb} \edb & \color{\nspab} \pab & }
			\bottomrule
		\end{tabular}
	\end{minipage}
\end{table*}

\subsection{Full-Orbit Thermal Measurement}
\label{sec:daynight}
The first way to resolve the degeneracy between Bond albedo and heat recirculation is by combining thermal eclipse and phase measurements to infer the planet's nightside effective temperature. This requires phase variations at thermal wavelengths, again defined as longward of $0.8~\mu$m. Such phase observations are more time-intensive than eclipses, and therefore less widely available. There are six planets with at least one published phase measurement: HD~149026b, HD~189733b, HD~209458b, WASP-12b, WASP-18b, and WASP-43b. As in Section~\ref{sec:dayEB}, we include band-integrated spectroscopy when it complements photometric observations \citep[only one case at the moment:][]{stevenson2014thermal}. For non-detections, an $n\sigma$ upper limit of $\alpha$ is assumed to have a value and uncertainty of $\alpha / 2$ and $\alpha / (2n)$ respectively. Table~\ref{tab:alt2edpvdata} shows the data for this sample.

Observational references are as follows: HD~149026b: \citet{stevenson2012transit}; \citet{knutson20098}; HD~189733b: \citet{evans2013deep}; \citet{crouzet2014water}; \citet{barnes2007limits}; \citet{knutson20123}; \citet{charbonneau2008broadband}; \citet{todorov2014updated}; \citet{agol2010climate}; \citet{deming2006strong}; \citet{knutson2009multiwavelength}; HD~209458b: \citet{rowe2008very}; \citet{richardson2003infrared}; \citet{knutson20083}; \citet{zellem20144}; \citet{cowan2007hot}; WASP-12b: \citet{fohring2013ultracam}; \citet{croll2014near}; \citet{crossfield2012re}; \citet{stevenson2014deciphering}; \citet{swain2013probing}; \citet{cowan2012thermal}; WASP-18b: \citet{maxted2013spitzer}; \citet{nymeyer2011spitzer}; WASP-43b: \citet{stevenson2014thermal}; \citet{wang2013ground}; \citet{zhou2014ks}; \citet{blecic2013spitzer}.

\begin{figure}
	\centering
	\includegraphics[width=1.0\linewidth]{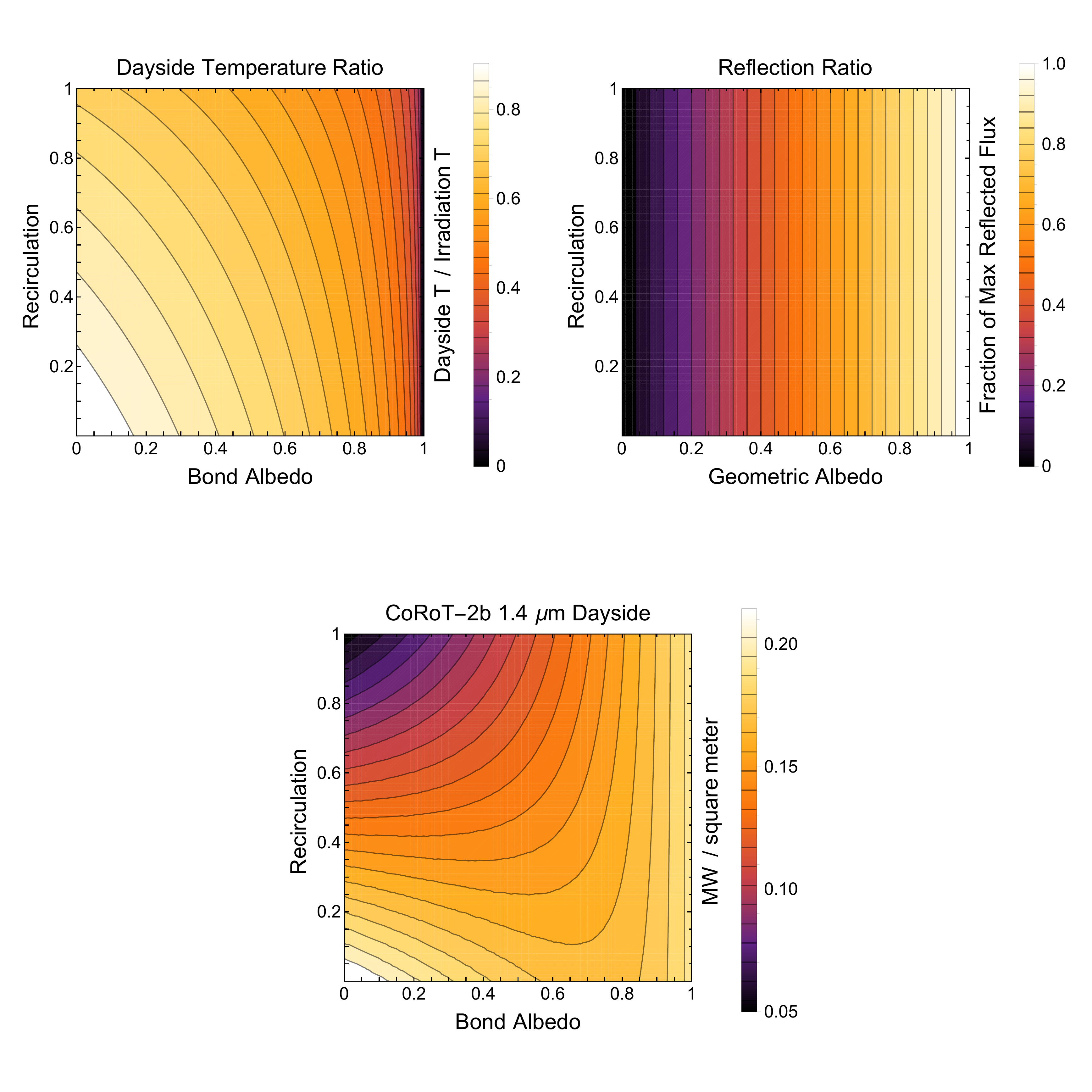}
	\caption{Dayside temperature (top left), reflection (top right), and dayside flux for CoRoT-2b at $1.4~\mu$m (bottom), shown as a function of albedo and recirculation (assuming the geometric and Bond albedos are equal to one another; see Section~\ref{sec:geotobond} for caveats.) At high albedo, the NIR dayside flux roughly parallels reflected starlight, while at low albedo, the dayside flux is mostly thermal emission and hence depends on day-night heat transport.}
	\label{fig:waveband_CoRoT2b_1_4_um}
\end{figure}

\subsubsection{Reflected Infrared Light}
\label{sec:reflectIR}
The light emanating from a planet's dayside is a combination of thermal emission and reflected starlight. We plot an example for CoRoT-2b at $1.4~\mu$m in the bottom panel of Figure~\ref{fig:waveband_CoRoT2b_1_4_um}, assuming the geometric and Bond albedos are equal (though as described in Section~\ref{sec:geotobond} the proper conversion is more involved.) If the planet has low albedo in this scenario, then the $1.4~\mu$m flux is almost entirely thermal emission and depends primarily on day--night heat transport. In the high albedo limit, on the other hand, the NIR flux is primarily reflected light and so varies linearly with the geometric albedo. In other words, even eclipse measurements at wavelengths \emph{greater} than $0.8~\mu$m are potentially contaminated by reflected starlight.

\begin{figure}
	\centering
	\includegraphics[width=1.0\linewidth]{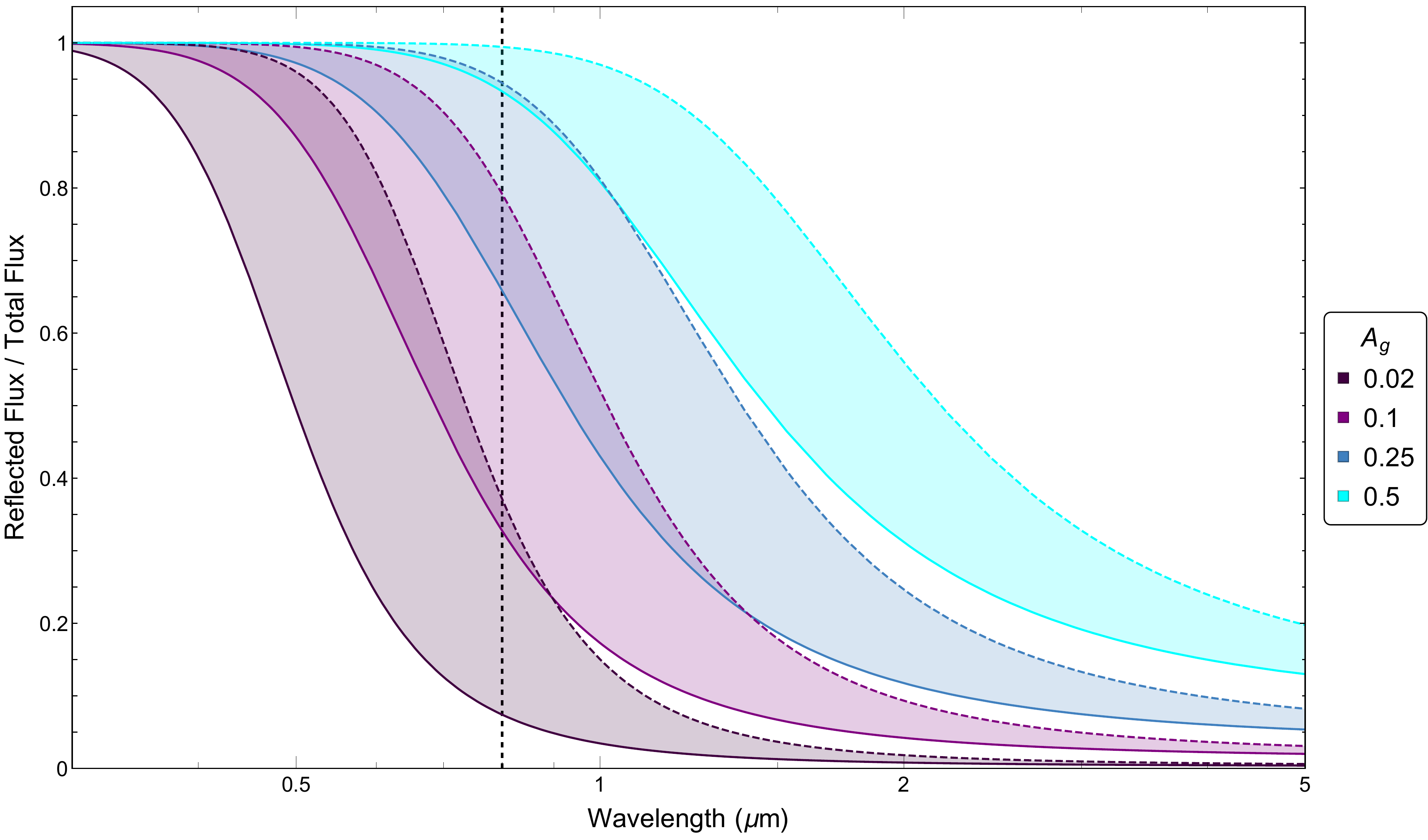}
	\caption{Contribution of planetary reflected light as a function of wavelength for different geometric albedos (lighter = higher), assuming blackbody emission and $q = \frac{5}{4}$ (see Section~\ref{sec:geotobond}). This example planetary system assumes $T_{*} = 6100~K$ and $a_{*} = 4.8$, both weighted means of the fifteen planets in Table~\ref{tab:alt2edpvdata}. Albedo regions are bounded by dayside temperature limits using Equation~\ref{eq:day}: solid lines denote no recirculation ($\varepsilon = 0$), dashed lines denote perfect recirculation ($\varepsilon = 1$). The vertical dashed line indicates our cutoff of $0.8~\mu$m between reflected light and thermal emission.}
	\label{fig:IRreflect}
\end{figure}

Furthermore, Figure~\ref{fig:IRreflect} shows the reflected light contribution to dayside flux as a function of wavelength for a hypothetical gray planet (with temperature limits derived from Equation~\ref{eq:day}; see Section~\ref{sec:daynightsigs}.) Reflected light dominates at ultraviolet wavelengths as expected, but its prevalence continues well through the near-infrared. For reasonable system parameters, reflected light contributes $\gtrsim 10$ per cent of the NIR flux (this reflected contribution goes up if $T_{*}$ is increased, or if $T_{d}$ or $a_{*}$ are decreased.) Even if molecular absorption depresses the reflectance in certain bands, it is likely that eclipse measurements in NIR water opacity windows ($J$, $H$, and $K$) are contaminated by reflected light at the $\gtrsim 10$ per cent level.

\subsubsection{Confidence Regions}
\label{sec:daynightsigs}
In estimating effective dayside and nightside temperatures for each planet, we use Monte Carlo simulations with $5000$ steps to propagate uncertainties. We assume uncertainties on observed quantities to be Gaussian and symmetrical; when asymmetrical uncertainty bars are reported, we adopt their mean. The planetary irradiation temperature is first computed as described in Section~\ref{sec:Teff}. Brightness temperatures are calculated for each appropriate measurement from Table~\ref{tab:alt2edpvdata} using Equation~\ref{eq:bright}, propagating uncertainty on stellar effective temperature, eclipse depth, phase amplitude, and transit depth. For observations \emph{not} listed in cyan (isolated eclipses, or partial phase curves), we conservatively inflate the published uncertainty by the factor $f_\text{sys} = 3$ \citep{hansen2014broadband}. Reflected light contributions are subtracted from all dayside eclipse depths---using planetary radius, $R_{p}$, and semi-major axis---assuming infrared geometric albedos to be normally distributed with mean 0.07 and width 0.01 (this is the distribution of uncorrected optical geometric albedo values described in Section~\ref{sec:geotobond}). In cases where a brightness temperature is calculated as $0~K$ for all MC steps, we assume $100~K$ uncertainty in subsequent propagations. We then compute the effective dayside and nightside temperatures using the EWM of the corresponding brightness temperatures (as this requires no a priori temperature assumption and produces similar values to the PWM.)

Our parameterization of recirculation neglects any treatment of poleward heat transport. The most extreme meridional temperature profiles are either uniform in the North-South direction (perfect poleward transport) or $T \propto \cos^{\frac{1}{4}}\theta$, where $\theta$ is latitude (no poleward transport). The difference in effective temperature seen by an equatorial observer is $\left( 1 / 4 \right)^{1/4} T_{0}$ versus $\left( 8 / 3\pi^{2} \right)^{1/4} T_{0}$, a 1 per cent discrepancy. We incorporate this worst-case systematic uncertainty in quadrature for all effective temperature estimates.

Understandably, the number of brightness temperature measurements at distinct wavelengths for a planet affects the accuracy of the effective temperature estimate. In a Monte Carlo analysis using J.J.~Fortney atmospheric models, \cite{cowan2011statistics} estimated systematic errors of 7.6 per cent in effective temperature when only a single observation was used (note that we only consider planets with at least two measurements), down to approximately 2.5 per cent for four or more measurements. We conservatively adopt a similar sliding scale of 8 per cent down to 3 per cent systematic uncertainty in effective temperature over the same observation number range, again added in quadrature.

Once we have dayside and nightside effective temperatures---and realistic uncertainties---for the six exoplanets, it is possible to infer each planet's Bond albedo and day--night heat transport efficiency using the parameterization of \cite{cowan2011statistics}:
\begin{equation}
\label{eq:day}
T_{d} = T_{0} (1 - A_{B})^{1/4} \left(\frac{2}{3} - \frac{5}{12} \varepsilon\right)^{1/4},
\end{equation}
and
\begin{equation}
\label{eq:night}
T_{n} = T_{0} (1 - A_{B})^{1/4} \left(\frac{\varepsilon}{4}\right)^{1/4},
\end{equation}
where both $ A_{B} $ and $ \varepsilon $ can take values between zero and unity.

We create $\chi^{2}$ surfaces for each planet based on our estimated dayside and nightside effective temperatures and using Equations~\ref{eq:day} and \ref{eq:night}. We calculate $\chi^{2}$ on a $101 \times 101$ grid in $A_{B}$ and $\varepsilon$, then interpolate the intermediate values. The $1\sigma$, $2\sigma$, and $3\sigma$ confidence intervals are defined as $\Delta\chi^{2} = \{1, 4, 9\}$ respectively above the minimum, $\chi_\text{min}^{2}$, where $\chi_\text{min}^{2} \approx 0$ for most planets because we have two constraints and two model parameters. Generating the $\chi^{2}$ surfaces involves numerical integrations of Planck functions, which can be computationally intensive. We therefore create a database of relevant integrals; with $10^{4}$ grid points tested per effective temperature, this database decreases computational time by more than 95 per cent.

\begin{figure}
	\centering
	\includegraphics[width=1.0\linewidth]{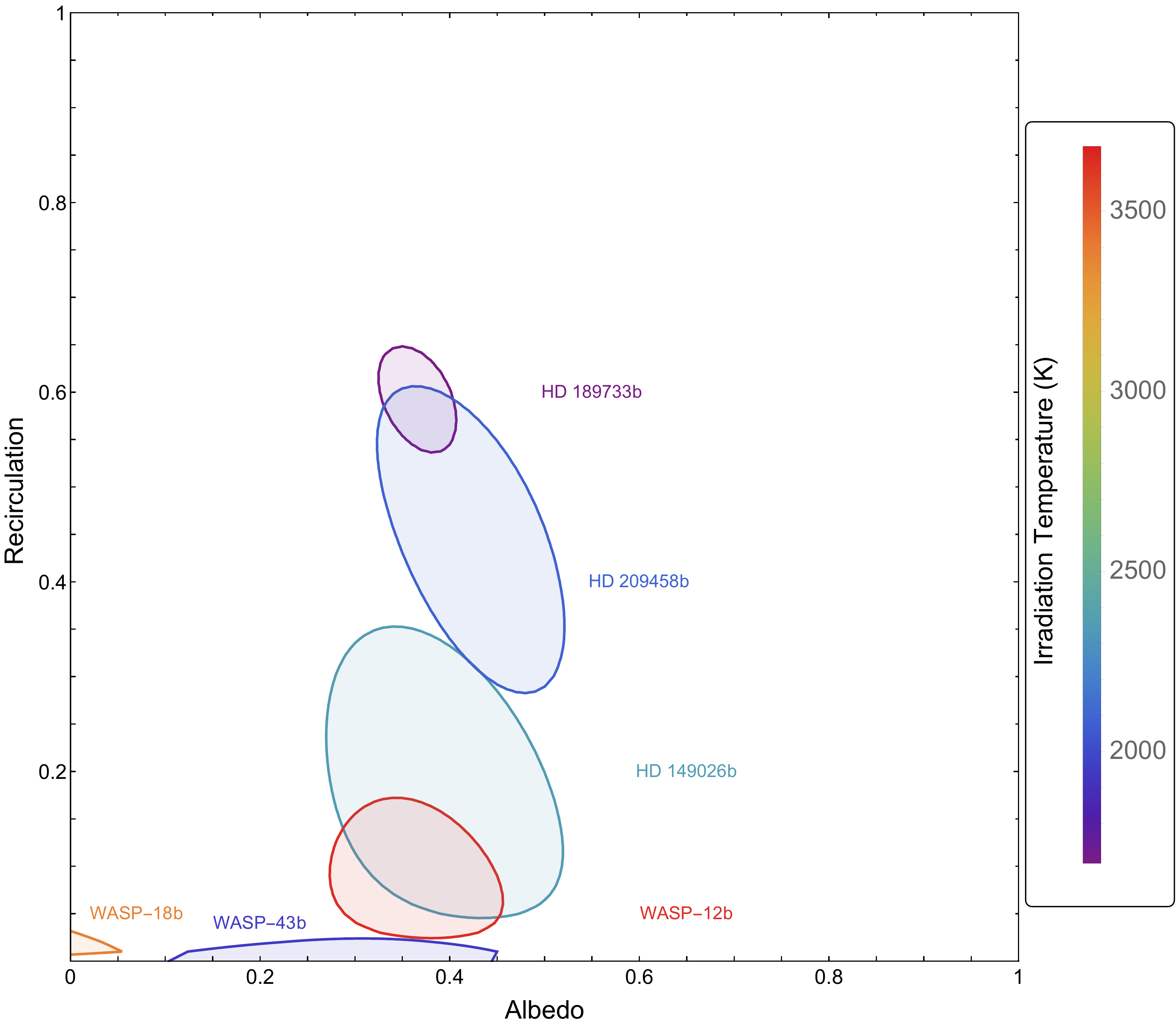}
	\caption{Composite $1\sigma$ confidence regions for thermal observation planets, as calculated from the error-weighted mean dayside and nightside brightness temperatures. Here the horizontal axis measures \emph{Bond} albedo. Bounding curve colors indicate irradiation temperature: red~=~warmer, purple~=~cooler. The inflationary factor $f_\text{sys} = 3$ is applied to infrared eclipse and phase uncertainties as noted in Table~\ref{tab:alt2edpvdata}, but adopting the published eclipse uncertainties barely modifies the confidence intervals.}
	\label{fig:bright_one_sigs_therm}
\end{figure}

We plot the $1\sigma$ confidence intervals for the six exoplanets with full-orbit thermal measurements in Figure~\ref{fig:bright_one_sigs_therm}. Each planet is colored according to irradiation temperature, essentially the incident stellar flux. Since these planets have benefited from intensive observational campaigns, omitting the $f_\text{sys} = 3$ uncertainty inflation produces nearly identical confidence intervals.

\subsection{Geometric Albedo Measurement}
\label{sec:dayalb}
The alternative approach to resolving the albedo versus heat-transport degeneracy of thermal eclipses is to also consider eclipse measurements at optical wavelengths. For our purposes, optical eclipses are those shortward of $0.8~\mu$m; these observations allow us to infer the planet's optical geometric albedo. Our literature review finds nine planets with published eclipse depths at thermal and optical wavelengths, but lacking infrared phase variations (Table~\ref{tab:alt2edpvdata}). Note that HD~189733b and HD~209458b benefit from \emph{both} infrared phases and visible eclipses.

Planets in this sample include CoRoT-1b \citep{alonso2009secondary1, zhao2012ground, gillon2009vlt, rogers2013benchmark, deming2011warm}, CoRoT-2b \citep{alonso2009secondary2, wilkins2014emergent, alonso2010ground, deming2011warm}, HAT-P-7b \citep{esteves2014changing, christiansen2010studying}, Kepler-5b \citep{esteves2014changing, desert2011atmospheres}, Kepler-6b \citep{esteves2014changing, desert2011atmospheres}, Kepler-7b \citep{esteves2014changing, demory2013inference}, Kepler-13Ab \citep{esteves2014changing, shporer2014atmospheric}, TrES-2b \citep{esteves2014changing, croll2010near, oDonovan2010detection}, and WASP-19b \citep{abe2013secondary, zhou2013examining, bean2013ground, zhou2014ks}.

\subsubsection{Thermal Contamination}
\label{sec:thermalcontam}
In order to extract a geometric albedo from an optical eclipse, we must correct the eclipse depth for thermal emission from the planet ``leaking" into the visible band \citep{cowan2011statistics, heng2013understanding}. In practice, one estimates the planet's dayside effective temperature and extrapolates this into the optical to account for thermal emission at visible wavelengths. However, this procedure is complicated by the fact that real hot Jupiters are vertically and horizontally inhomogeneous, so they emit at higher brightness temperatures in the optical than in the thermal infrared.

\begin{figure}
	\centering
	\includegraphics[width=1.0\linewidth]{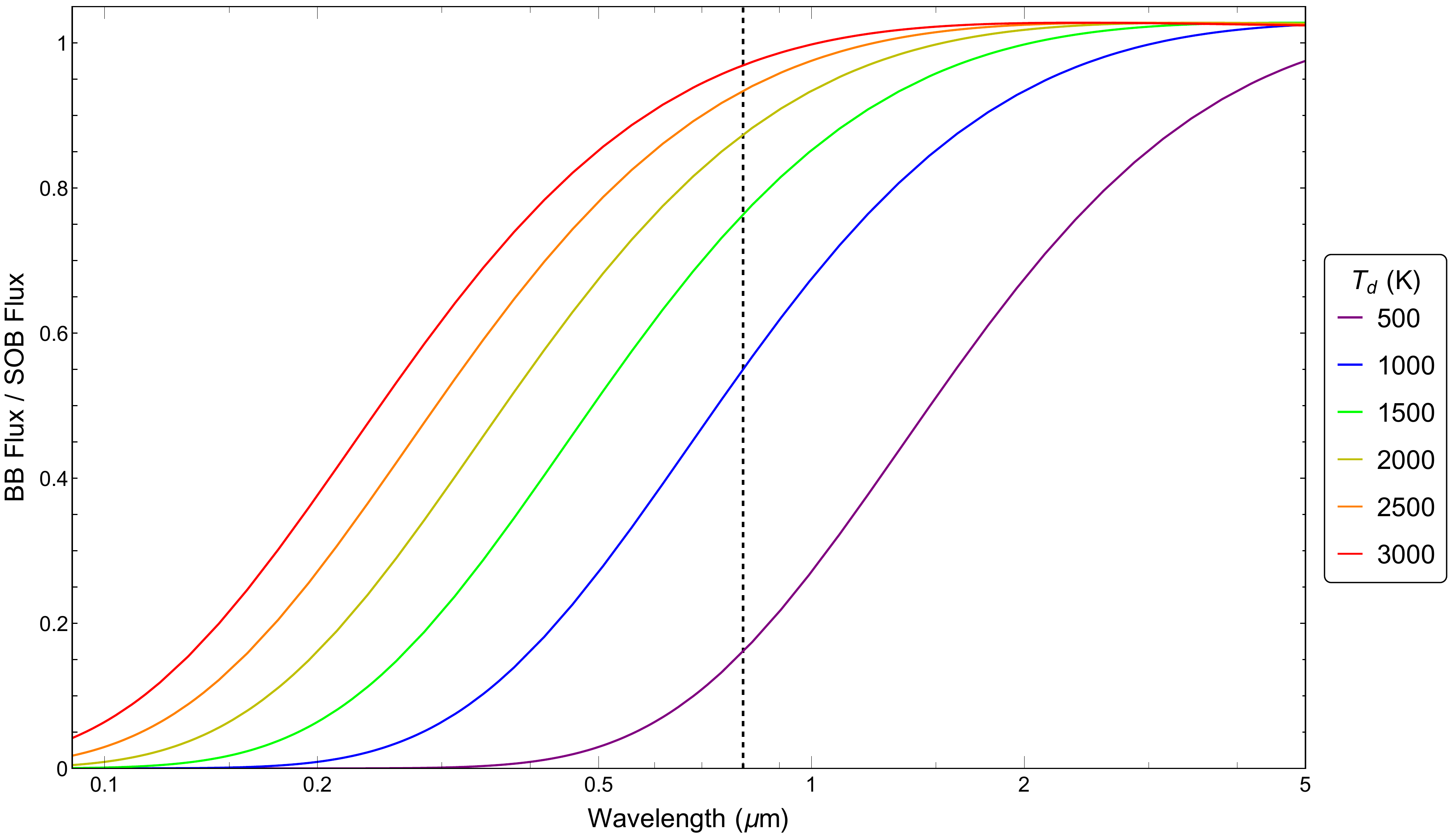}
	\caption{Flux ratio between blackbody and ``Sum of Blackbodies" spectra for various effective dayside temperatures in $500~K$ increments. Curves are colored according to temperature: red~=~warmer, purple~=~cooler. The vertical dashed line indicates our chosen threshold wavelength, $0.8~\mu$m, between reflected light and thermal emission.}
	\label{fig:BBSoBflux}
\end{figure}

 Even if every location on a planet emits as a blackbody (BB), the resulting spectrum will not be a Planck curve. For a planet in the zero-albedo and zero-recirculation limit, the equilibrium temperature at any dayside location is described by $T = T_{0} \cos^{\frac{1}{4}}\gamma$, where $\gamma$ is the angle from the sub-stellar point ($\gamma= \frac{\pi}{2}$ at the terminator.) Each annulus of the dayside thus radiates at a different blackbody temperature, and together they produce a ``Sum of Blackbodies" (SoB) spectrum \citep[this is analogous to the multicolor blackbody spectra used to model accretion disks;][]{Mitsuda_1984}. For fixed bolometric flux, BB and SoB spectra produce nearly identical flux at thermal wavelengths: the SoB is 1--2 per cent fainter than the BB. At optical wavelengths, however, a BB spectrum underestimates the flux by a factor of a few, as seen in Figure~\ref{fig:BBSoBflux}.
 
 Moreover, the optical photosphere should be deeper and hotter than the infrared photosphere, in a cloud-free atmosphere \citep{allard2001limiting, fortney2008unified, cowan2011model}. The combination of these two effects is that a na\"ive blackbody extrapolation from the infrared to the optical may underestimate thermal emission by a factor of 3--10. In other words, while the hottest planets have the greatest thermal emission at optical wavelengths, somewhat cooler planets have optical emission that is harder to estimate.

Once an optical eclipse has been corrected for thermal contamination, the geometric albedo can be calculated using
\begin{equation}
\label{eq:ageo}
A_{g} = \delta_\text{ecl}^\text{ref} \left(\frac{a}{R_{p}}\right)^{2},
\end{equation}
where $\delta_\text{ecl}^\text{ref}$ is the \emph{reflected light} eclipse depth. Note that geometric albedo is a function of wavelength, while hot Jupiter eclipse depths have typically only been measured in a single optical broadband.

\begin{figure}
	\centering
	\includegraphics[width=1.0\linewidth]{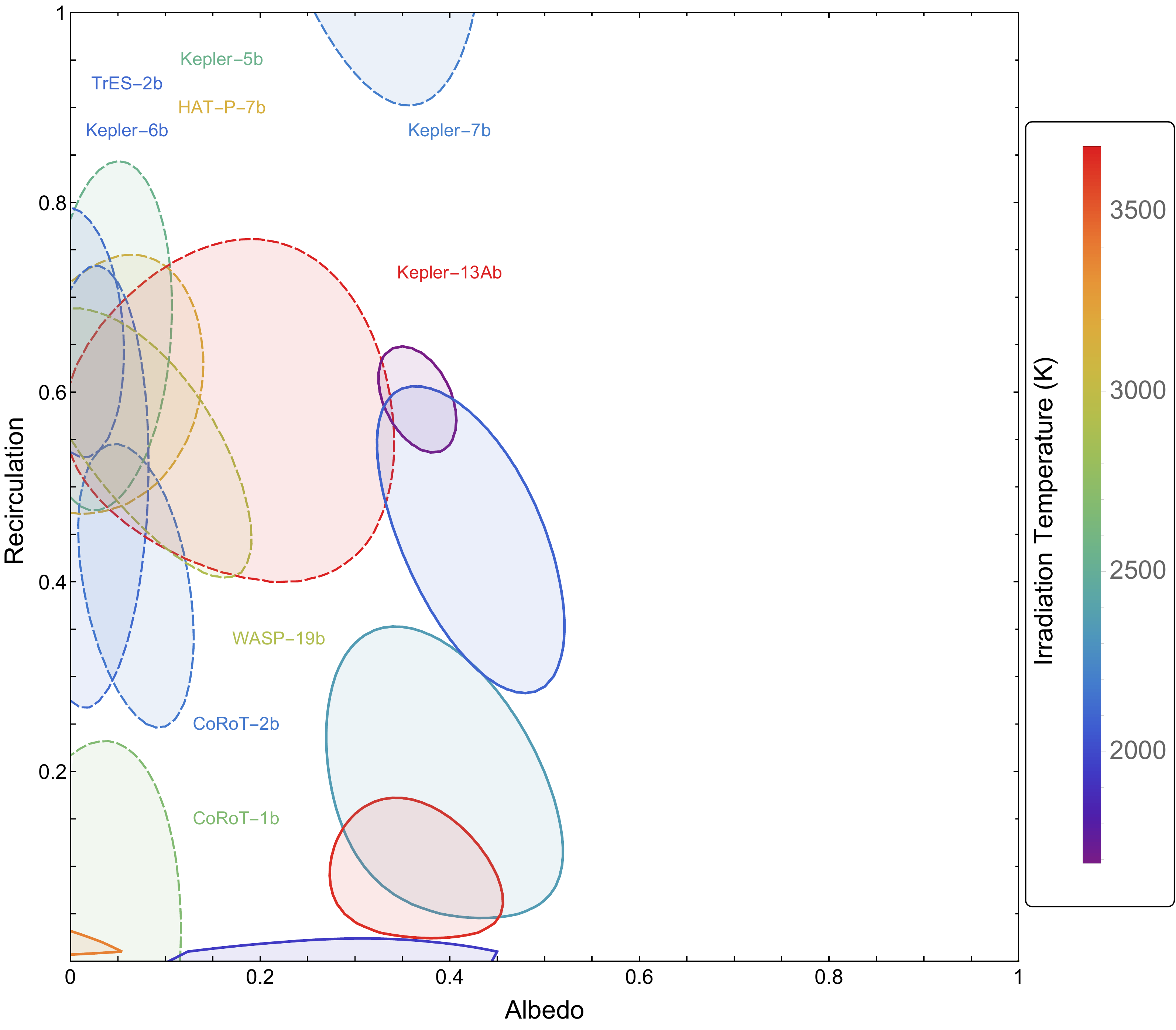}
	\caption{Composite $1\sigma$ confidence regions for eclipse-only planets as calculated from brightness temperatures (using EWM) and geometric albedos, shown with thermal observation planets for comparison. Here the horizontal axis measures different quantities: \emph{geometric} albedo at visible wavelengths for eclipse-only planets (dashed lines), \emph{Bond} albedo for thermal observation planets (solid lines). Bounding curve color follows Figure~\ref{fig:bright_one_sigs_therm}. The inflationary factor $f_\text{sys} = 3$ is applied to infrared eclipse and phase uncertainties as noted in Table~\ref{tab:alt2edpvdata}. Adopting published uncertainties across the board results in similar confidence intervals.}
	\label{fig:bright_one_sigs_fsys}
\end{figure}

\subsubsection{Confidence Regions}
\label{sec:dayalbsigs}
Our dayside temperature analysis for planets in the eclipse-only sample is analogous to Section~\ref{sec:daynightsigs}, and we again perform Monte Carlo simulation with $5000$ steps for all uncertainty propagation. For optical eclipses, we optimize computation by first calculating the \emph{uncertainty} of each thermally-corrected geometric albedo, using Equation~\ref{eq:ageo} and propagating uncertainties in EWM dayside effective temperature, stellar effective temperature, transit depth, eclipse depth, planetary radius, and semi-major axis. Our thermal correction uses equal contributions of BB and SoB spectra, assuming 10 per cent increase in dayside temperature to account for the vertical temperature profile effect noted in Section~\ref{sec:thermalcontam}. To acknowledge variability with this effect, we add a 5 per cent systematic uncertainty in quadrature to the calculated geometric albedo uncertainty. We then construct $\chi^{2}$ surfaces for each planet as described in Section~\ref{sec:daynightsigs}. For optical measurements, we recompute our thermal contamination correction at each $\chi^{2}$ grid point to determine specific \emph{values} of geometric albedo (since the underlying dayside temperature varies in the albedo--recirculation plane.) Note that HD~189733b has two distinct optical eclipses; we use the weighted mean of both corrected geometric albedos in our $\chi^{2}$ calculations for this planet.

\begin{table*}
	\begin{minipage}{95mm}
		\centering
		\caption{Resulting parameters for all planets as calculated from brightness temperatures (via EWM) and geometric albedos, assuming applicable $f_\text{sys} = 3$ uncertainty inflation. Thermal observation planets are listed first (with \emph{Bond} albedos), followed by eclipse-only planets (with \emph{geometric} albedos at visible wavelengths.) Low and high values are obtained from the confidence regions of Figures~\ref{fig:bright_one_sigs_therm} and \ref{fig:bright_one_sigs_fsys}; fit values are taken to be the grid location of $\chi_\text{min}^{2}$.}
		\label{tab:fitval}
		\begin{tabular}{c c c c c c c c}
			\toprule
			\multirow{2}*{Planet} & \multicolumn{3}{c}{Albedo} & \multicolumn{3}{c}{Recirculation} & \multirow{2}*{$\chi_\text{min}^{2}$}\\
			& Low & \textbf{Fit} & High & Low & \textbf{Fit} & High & \\
			\midrule
			\begin{filecontents*}{table3.csv}
exo,fitbo,lbo,hbo,fitep,lep,hep,minchi,n
HD 149026b,0.41,0.27,0.519,0.15,0.045,0.353,0.0005,2
HD 189733b,0.37,0.325,0.407,0.59,0.536,0.648,0.0051,2
HD 209458b,0.43,0.323,0.521,0.44,0.283,0.606,0.0013,2
WASP-12b,0.37,0.273,0.456,0.07,0.024,0.172,0.0028,2
WASP-18b,0,0,0.054,0.01,0.007,0.032,2.2571,2
WASP-43b,0.29,0.104,0.45,0,0,0.024,0.7581,2
,,,,,,,,
CoRoT-1b,0.01,0,0.117,0,0,0.232,0.0002,2
CoRoT-2b,0.07,0.008,0.13,0.4,0.246,0.545,0.0005,2
HAT-P-7b,0.04,0,0.14,0.61,0.472,0.745,0.0012,2
Kepler-5b,0.04,0,0.107,0.67,0.476,0.844,0.001,2
Kepler-6b,0.02,0,0.082,0.51,0.267,0.733,0.005,2
Kepler-7b,0.34,0.258,0.426,1,0.902,1,0.9611,2
Kepler-13Ab,0.18,0,0.341,0.58,0.4,0.761,0.0004,2
TrES-2b,0.01,0,0.056,0.67,0.532,0.795,0.0097,2
WASP-19b,0.08,0,0.191,0.56,0.404,0.688,0.0009,2
\end{filecontents*}
\csvreader[head to column names, late after line=\\]{table3.csv}{}{\exo & \lbo & \bfseries \fitbo & \hbo & \lep & \bfseries \fitep & \hep & \minchi}
			\bottomrule
		\end{tabular}
	\end{minipage}
\end{table*}

In Figure~\ref{fig:bright_one_sigs_fsys} we compare the $1\sigma$ confidence intervals of the nine eclipse-only planets to those of the six planets with full-orbit thermal observations. Note that for optical eclipses we implicitly assume $A_{B} = A_{g}$, but the actual conversion between geometric and Bond albedo is more complicated (Section~\ref{sec:geotobond}). Regions are again colored by irradiation temperature, while sample group is denoted by the line style of bounding curve. Though the $f_\text{sys}$ uncertainty inflation is included for isolated thermal eclipses, taking the published uncertainties at face value produces nearly identical confidence intervals.

Based solely on dayside effective temperatures (Figure~\ref{fig:TdToscatter}), one might conclude that all planets have roughly the same Bond albedo and heat transport efficiency. Figure~\ref{fig:bright_one_sigs_fsys} dispels this notion at high significance. The $1\sigma$ intervals for all the planets in Figure~\ref{fig:bright_one_sigs_fsys} are listed in Table~\ref{tab:fitval}: thermal observation planets first, eclipse-only planets second. All best-fit parameters are defined as the location of $\chi_\text{min}^{2}$ on the computed grids.

\section{Discussion}
\label{sec:discuss}

\subsection{Sources of Error and Uncertainty}
\label{sec:errunc}
It is worth summarizing the various sources of uncertainty and error that we account for in order to produce Figure~\ref{fig:bright_one_sigs_fsys}. For thermal measurements, we first compute brightness temperatures, accounting for the uncertainties on eclipse depth \citep[inflated by a factor of 3 if based on a single occultation;][]{hansen2014broadband}, transit depth, and stellar effective temperature, and also compute each planet's irradiation temperature, accounting for uncertainty in stellar effective temperature and scaled semi-major axis. At this stage, we also account for reflected light contamination (non-zero near-infrared geometric albedo). Nightside brightness temperatures are derived the same way, but additionally depend on the thermal phase amplitude and its uncertainty. In converting brightness temperatures to effective temperatures, we account for unknown meridional heat transport and incomplete spectral coverage. The conversion from dayside brightness temperatures to dayside effective temperature is reasonable for hot Jupiters because of their relatively isothermal vertical structure; the conversion may be more fraught for the nightsides of hot Jupiters.

For optical eclipses, we first correct eclipse depths for thermal contamination, accounting for uncertainty on dayside temperature, transit depth, and stellar effective temperature. In this process we also account for vertical and horizontal temperature profiles that conspire to increase optical thermal emission. We next convert the reflected light eclipse depth to an optical geometric albedo, accounting for uncertainties in eclipse depth, planetary radius, and scaled semi-major axis. We assume $A_{B} = A_{g}$ for the purposes of constraining heat transport in Figure~\ref{fig:bright_one_sigs_fsys} (see Section~\ref{sec:geotobond} for caveats), but this assumption in no way affects our inferred geometric albedo for these planets.

Crucially, for every ``correction'' that we apply, we add appropriate uncertainty in our inference, either by randomly varying parameters in the Monte Carlo, or by adding systematic uncertainty in quadrature to the formal errors. Our inferences of heat transport, Bond albedo, and geometric albedo are therefore conservative.

\subsection{Dayside Temperatures}
\label{sec:Tdtrend}
The upward trend in dayside effective temperature with irradiation temperature in Figure~\ref{fig:TdToscatter} is unsurprising: one expects highly-irradiated planets to be hotter. The black lines in the plot can be thought of as limiting cases of either heat recirculation or Bond albedo. In the zero-albedo limit, the black lines correspond to $\varepsilon = 0$ (solid), $\varepsilon = 0.4$ (dashed), and $\varepsilon = 1$ (dotted). Alternatively, in the zero-recirculation limit, the black lines correspond to $A_{B} = 0$ (solid), $A_{B} = 0.25$ (dashed), and $A_{B} = 0.625$ (dotted). Therefore, planets that lie above the solid black line must have an internal energy source, while planets lying below the dotted black line must have non-zero Bond albedo.

We can also consider the qualitative claim from \citet{cowan2011statistics} that $T_{d}$ increases \emph{disproportionately} with $T_{0}$. We ignore planets with significantly eccentric orbits, as this complicates their energy budget (these planets are denoted in red in Figure~\ref{fig:TdToscatter}.) With double the planets and more data per planet, we find that $T_{d}~=~\num[scientific-notation=false]{-90 \pm 80} + \num[scientific-notation=false]{0.87 \pm 0.05} T_{0}$. The $\chi^{2}$ per datum of the fitted trend is $1.4 \pm 0.4$, which is a reasonable fit. This is consistent with---but does not strengthen---the claim that planets receiving more stellar flux generally have lower Bond albedo and/or less efficient heat transport.

\begin{table*}
	\begin{minipage}{155mm}
		\centering
		\caption{Geometric albedo values, optical waveband starlight fraction, and Bond albedo limits (using Equations~\ref{eq:ABmin}--\ref{eq:ABhigh}) for planets with visual eclipses. Observations are denoted by central wavelength and bandwidth. Geometric albedo values are considered as follows: ``uncorrected" uses Equation~\ref{eq:ageo} assuming no thermal contamination; ``simple correction" accounts for thermal emission with a BB spectrum at our error-weighted mean effective dayside temperature; ``full correction" uses the method described in Section~\ref{sec:dayalbsigs} (equal BB + SoB contributions with vertical temperature profile effect), again using our EWM dayside temperature. Uncertainties on starlight fraction are negligible and thus omitted. All Bond albedo calculations use the ``full correction" values and assume the phase integral factor $q~=~\frac{5}{4}$.}
		\label{tab:whitegrayblack}
		\begin{tabular}{c S[scientific-notation=false, table-parse-only] S[scientific-notation=false, table-format=1.3(3)] S[scientific-notation=false, table-format=1.3(3)] S[scientific-notation=false, table-format=1.3(3)] S[scientific-notation=false, table-format=1.3] S[scientific-notation=false, table-format=1.3] S[scientific-notation=false, table-format=1.3] S[scientific-notation=false, table-format=1.3] l@{}}
			\toprule
			\multirow{2}*{Planet} & \multirow{2}*{Wavelength ($\mu$m)} & \multicolumn{3}{c}{Geometric Albedo} & \multicolumn{1}{c}{\multirow{2}*{$f_{*}^\text{opt}$}} & \multicolumn{3}{c}{Bond Albedo} & \\
			& & {Uncorrected} & {Simple Correction} & {Full Correction} & & {Min} & {Gray} & {High} & \\
			\midrule
			\begin{filecontents*}{table4.csv}
exo,wave,band,agno,aguno,agbb,agubb,ag,agu,flux,funxu,bgb,bg,bgw
CoRoT-1b,0.6,0.42,0.213,0.087,0.129,0.082,0.043,0.077,0.485,0.005,0.026,0.053,0.284
CoRoT-2b,0.6,0.42,0.101,0.035,0.09,0.034,0.069,0.06,0.472,0.006,0.04,0.086,0.305
HAT-P-7b,0.65,0.4,0.262,0.06,0.156,0.054,0.051,0.073,0.438,0,0.028,0.064,0.309
HD 189733b,0.37,0.16,0.374,0.113,0.374,0.113,0.374,0.124,0.091,0.003,0.043,0.468,0.497
,0.51,0.12,0.042,0.06,0.043,0.061,0.043,0.079,0.123,0.002,0.007,0.054,0.445
HD 209458b,0.5,0.3,0.039,0.037,0.039,0.036,0.039,0.062,0.384,0.006,0.019,0.049,0.327
Kepler-5b,0.65,0.4,0.107,0.022,0.079,0.025,0.04,0.058,0.439,0.001,0.022,0.05,0.302
Kepler-6b,0.65,0.4,0.06,0.022,0.047,0.023,0.025,0.055,0.436,0.001,0.014,0.031,0.296
Kepler-7b,0.65,0.4,0.314,0.073,0.315,0.072,0.313,0.088,0.44,0,0.172,0.392,0.452
Kepler-13Ab,0.65,0.4,0.468,0.031,0.318,0.067,0.175,0.113,0.405,0.009,0.088,0.218,0.386
TrES-2b,0.65,0.4,0.03,0.007,0.022,0.008,0.007,0.051,0.439,0,0.004,0.008,0.284
WASP-19b,0.685,0.53,0.248,0.12,0.177,0.113,0.099,0.108,0.537,0.003,0.067,0.124,0.298
\end{filecontents*}
\csvreader[head to column names, late after line=\\]{table4.csv}{}{\exo & \wave \pm \band & \agno \pm \aguno & \agbb \pm \agubb & \ag \pm \agu & \flux & \bgb & \bg & \bgw & }
			\bottomrule
		\end{tabular}
	\end{minipage}
\end{table*}

\subsection{Thermal Phase Measurements}
\label{sec:thermalobs}
Figure~\ref{fig:bright_one_sigs_therm} shows a tendency towards lower recirculation efficiency as irradiation increases, in agreement with previous findings \citep{cowan2011statistics, cowan2012thermal, perez2013atmospheric}. The irradiation temperatures of these planets span approximately $2000~K$, corresponding to $\varepsilon = 0.59$ for HD~189733b at the cool end ($T_{0} = 1695~K$) and $\varepsilon = 0.01$ for WASP-18b ($T_{0} = 3387~K$). The irradiation of WASP-12b is actually $\sim 260~K$ higher than WASP-18b, but their recirculation probability distribution functions overlap (Table~\ref{tab:fitval}).

The notable exception to the $T_{0}$--$\varepsilon$ trend is WASP-43b, with $T_{0} = 1943~K$ but exhibiting virtually no heat transport ($\varepsilon = 0$ with $\chi_\text{min}^2 = 0.758$). Our recirculation value is in agreement with the redistribution factor of \citet{stevenson2014thermal}, and our best-fit Bond albedo ($A_{B} = 0.29$) is also consistent at the $1\sigma$ level. These parameters translate into a cold nightside temperature (nominally $T_{n} \lesssim 465~K$), which we routinely find to be consistent with zero. Coupled hydrodynamic and radiative transfer simulations of this planet were able to reproduce its dayside---but not nightside---emission \citep{kataria2014circulation}, so the poor heat transport of this planet remains a mystery.

The thermal measurements for WASP-18b suggest $A_{B}~\lesssim~0.05$ at $1\sigma$. The planet's best-fit parameters would lie outside the plot to the left, which is indicative of either an internal heat source (identical to a negative Bond albedo in our parametrization) or underestimated observational uncertainties. Kepler-7b and WASP-43b also have $\chi_\text{min}^{2}$ well above zero, suggesting that either our model assumptions or the published uncertainties are incorrect. Each of these planets would benefit from more thermal eclipse and phase measurements in order to reduce the systematic uncertainty in dayside and nightside effective temperatures.

HD~189733b and HD~209458b benefit from full-orbit thermal observations as well as optical eclipse measurements, allowing us to compare infrared-based Bond albedos to their optical geometric albedos. For HD~189733b we derive corrected geometric albedos of \num[scientific-notation=false]{0.37\pm0.12} at $0.37~\mu$m and \num[scientific-notation=false]{0.04\pm0.08} at $0.51~\mu$m, in agreement with \citet{evans2013deep}. This red-optical geometric albedo is much lower than our Bond albedo estimate of $[0.33, 0.41]$. For HD~209458b we obtain a corrected geometric albedo of \num[scientific-notation=false]{0.04\pm0.06} at $0.5~\mu$m, which agrees with \citet{rowe2008very}. However, our $1\sigma$ interval for Bond albedo is $[0.32, 0.52]$. Therefore, the tentative conclusion based on these two planets is that their Bond albedos are considerably higher than their optical geometric albedos.

\subsection{Eclipse-Only Measurements}
\label{sec:eclipseonly}
Most of the eclipse-only planets have low geometric albedos: $A_{g}~\lesssim~0.2$ (Figure~\ref{fig:bright_one_sigs_fsys}). As anticipated, Kepler-7b lies completely above this range \citep{demory2011high}, while the confidence region for Kepler-13Ab extends to a geometric albedo of 0.34. The nine planets exhibit a wide variety of recirculation efficiencies, from CoRoT-1b ($\varepsilon \approx 0.1$) to Kepler-7b ($\varepsilon\approx 0.95$). Eclipse-only planets with similar irradiation temperatures are found at different locations on the $\varepsilon$-axis, and we do not see evidence for a trend between irradiation temperature and recirculation efficiency as with the thermal observation planets. This is not surprising, since dayside measurements offer minimal leverage for inferring the nightside temperature. Strictly speaking we only include these planets in Figure~\ref{fig:bright_one_sigs_fsys} by assuming that $A_{B} = A_{g}$; the actual comparison is more complex (Section~\ref{sec:geotobond}).

Geometric albedo analyses encompassing several planets from our sample have been previously conducted. We compare our results in Table~\ref{tab:whitegrayblack} to overlapping planets from \citet{heng2013understanding}: HAT-P-7b, Kepler-5b, Kepler-6b, Kepler-7b, and TrES-2b. Our ``uncorrected" geometric albedos for all five planets show agreement within the stated confidence intervals. \citet{esteves2014changing} and \citet{angerhausen2014comprehensive} also consider these planets, in addition to Kepler-13Ab, under both zero and perfect heat redistribution. Our ``full correction" geometric albedos for five of the six planets are in agreement with values from both studies obtained in the maximum equilibrium temperature hypothesis (i.e. hotter dayside temperatures implying greater thermal contamination of the optical eclipse). However, we find Kepler-13Ab to have dissimilar geometric albedo when using stellar parameters from \citet{shporer2014atmospheric} with our greater thermal correction: \num[scientific-notation=false]{0.175 \pm 0.113} versus \num[scientific-notation=false]{0.404 \pm 0.055} and $\simeq 0$, respectively. Note this includes stellar effective temperature readjustment of Kepler-13A to $7650 \pm 250~K$, down from $8500 \pm 400~K$ \citep{szabo2011asymmetric}.

\subsection{Comparing Geometric to Bond Albedo}
\label{sec:geotobond}
In principle, an optical eclipse depth is related to a planet's Bond albedo \citep{rowe2006upper}. Indeed, for a range of Solar System planets and moons, the Bond albedo is roughly equal to the optical geometric albedo, albeit with a scatter of $\pm 30$ per cent. Given the possibility of inhomogeneous albedo, uncertainty in the scattering phase function, and unknown reflectance spectrum, it would be imprudent to extrapolate this trend to hot Jupiters. Moreover, \citet{marley1999reflected} demonstrated that simply varying the incident stellar radiation can alter Bond albedo by a factor of four for identical planets.

Our analysis of hot Jupiters suggests that their optical geometric albedos are systematically lower than their Bond albedos. There are three possible explanations for this discrepancy: (1) we have over-corrected the thermal contamination at optical wavelengths, (2) we have systematically underestimated the effective temperatures for planets with full-orbit phase variations, or (3) the geometric albedos of hot Jupiters are, in fact, systematically lower than their Bond albedos because of unexpected scattering phase functions and/or reflectance spectra.

We address the first hypothesis by listing three different geometric albedo calculations in Table~\ref{tab:whitegrayblack}: these differ in their treatment of optical thermal emission. For six of the eleven planets, there is little difference between the geometric albedo estimate after a simple blackbody subtraction as opposed to the scenario with higher optical brightness temperature. For the remaining five planets, the ``full correction" geometric albedos are lower than their ``simple correction" counterparts. The planets for which the details of thermal emission correction are more important tend to either have higher irradiation temperatures and so greater likelihood for unattributed thermal contamination in the optical (e.g. HAT-P-7b), or have precise optical eclipse measurements where minor changes to the dayside emission have a larger impact on constraining reflected light (e.g. TrES-2b). Our contamination analysis is also largely consistent with the geometric albedos inferred from higher equilibrium temperatures in both \citet{esteves2014changing} and \citet{angerhausen2014comprehensive}. Even in the unlikely event that hot Jupiters have optical dayside brightness temperatures equal to that in the mid-infrared, the optical geometric albedos for most planets are significantly lower than the Bond albedos inferred from thermal phase measurements.

The second solution to the geometric versus Bond albedo discrepancy is that hot Jupiters are much brighter in the NIR, and hence thermal phase measurements---mostly obtained with \emph{Spitzer} in the mid-IR---will underestimate their global temperatures and over-estimate their Bond albedos. However, neither the dayside emission spectra of individual planets, nor their aggregate spectrum, show strong broadband molecular features. This means that dayside brightness temperatures from the near- through mid-IR should be reasonable proxies for their effective temperatures. One may worry that flux is escaping the nightsides of hot Jupiters in the NIR, leading us to underestimate nightside bolometric flux, but that is ruled out for WASP-43b by HST/WFC3 phase measurements, which show \emph{no} nightside flux in the NIR \citep{stevenson2014thermal}. It would be useful to have full-orbit NIR phase curves of more planets in order to further test this hypothesis.

This leaves us with the third hypothesis, namely that the Bond albedos of most hot Jupiters are high, despite their low geometric albedos. The geometric albedo of a planet (light reflected back towards the illuminating star) is related to its spherical albedo (light reflected in all directions) by a phase integral, $A_{s} = q A_{g}$. Lambertian (diffuse) reflection results in $q= \frac{3}{2}$, while pure Rayleigh scattering produces $q = \frac{4}{3}$. In general, planets with atmospheres---including simulated hot Jupiters---have $1.0 < q < 1.5$ \citep{pollack1986planetesimal, burrows2010giant}. It would be useful to empirically constrain the scattering phase functions of hot Jupiters using data from space-based photometric missions. In the few cases where reflected phase variations have been measured, the spherical albedo appears so inhomogeneous that it is impossible to infer the phase-dependence of scattering \citep{demory2013inference, esteves2014changing}. If we assume that hot Jupiters are diffusely reflecting, then they have typical optical spherical albedos of 15 per cent, still well below the inferred Bond albedos.

The spherical albedo is related to the Bond albedo via a flux-weighted integral \citep{burrows2010giant}:
\begin{equation}
\label{eq:waveint}
A_{B} = \frac{\int_{0}^{\infty} A_{s}(\lambda) I_\text{inc}\,\text{d}\lambda}{\int_{0}^{\infty} I_\text{inc}\,\text{d}\lambda},
\end{equation}
where $I_\text{inc}$ is the SED of the incident stellar flux. The degree to which the optical spherical albedo impacts the Bond albedo depends on $f_{*}^\text{opt}$, the percentage of starlight emitted in the observed optical waveband, assuming blackbody radiation at $T_{*}$ (values of $f_{*}^\text{opt}$ are listed in Table~\ref{tab:whitegrayblack}). We consider limiting cases of Equation~\ref{eq:waveint}, assuming out-of-band wavelengths have average spherical albedos equal to $0$, the optical $A_{s}$, or $0.5$ respectively:
\begin{equation}
\label{eq:ABmin}
A_{B}^\text{min} = A_{s} f_{*}^\text{opt},
\end{equation}
\begin{equation}
\label{eq:ABgray}
A_{B}^\text{gray} = A_{s},
\end{equation}
\begin{equation}
\label{eq:ABhigh}
A_{B}^\text{high} = A_{s} f_{*}^\text{opt} + 0.5 (1-f_{*}^\text{opt}).
\end{equation}
Our limiting Bond albedos are summarized in Table \ref{tab:whitegrayblack}. In principle the out-of-band spherical albedo could be unity, but as this would result in Bond albedos so great that the planets would be cooler than is observed in the mid-IR, we adopt a more modest upper threshold in Equation~\ref{eq:ABhigh}.

The $A_{B}^\text{high}$ scenario is a reasonable match to the Bond albedos inferred from full-orbit thermal measurements. This suggests that most hot Jupiters have geometric albedos of $\approx 50$ per cent in the NIR and $\lesssim 10$ per cent in the optical. If hot Jupiters are Lambertian reflectors, the NIR/optical contrast is somewhat less severe. This scenario similar in spirit to the high Bond albedo combined with low red--NIR geometric albedo one can obtain with Rayleigh scattering \citep{marley1999reflected}, but with the opposite color. Note that the high infrared albedos we are hypothesizing contradict the low infrared geometric albedo we assumed when estimating reflected IR light in Section~\ref{sec:daynightsigs}. Using $A_{g}^\text{IR} = 0.5$ in our MC implies greater NIR contamination from reflected starlight, and hence lower dayside thermal flux with greater Bond albedo. The most extreme change is a 20 per cent increase in the Bond albedo of WASP-12b, but nonetheless our conclusions remain unaffected.

It is worth mentioning that geometric albedos of 60 per cent---from the optical through the NIR---were predicted for the hottest giant exoplanets due to reflective silicate clouds \citep[][]{sudarsky2000albedo}. In order to explain the low optical geometric albedo, one could invoke an optical absorber at low pressures, above the purported cloud deck. Such optical absorbers, originally theorized to explain hot Jupiter temperature inversions, could include gaseous TiO/VO \citep{fortney2007planetary} or S$_2$/HS \citep{zahnle2009atmospheric}. In this scenario, Kepler-7b would be unique not because of its clouds, but due to its dearth of optical absorbers.

Alternatively, since cloud reflection is a multiple-scattering process, single-scattering albedos even marginally below unity can result in a heavily muted geometric albedo \citep{dlugach1974optical, hu2014ammonia}. One might therefore explain the unusually red reflectance spectrum of hot Jupiters with a single cloud deck where individual cloud grains have nearly gray albedo.

Such scenarios might be tested by measuring the optical--infrared transit spectra of hot Jupiters. If the purported absorbers are located at sufficiently low pressures, and if the upper atmospheres of these planets are not too hazy \citep[][]{pont2013prevalence,gibson2013optical}, then we would expect larger effective radii in the optical than the infrared. If instead the red reflectance spectrum is due to multiple scattering within a single cloud deck, then the transit spectrum should be flat.

Moreover, the best way to investigate these hypotheses would be to obtain thermal phase measurements for the planets that have precise optical geometric albedo constraints, and vice versa.

\section*{Acknowledgments}
JCS was funded by an NSF GK-12 fellowship. Thanks to Nikole Lewis (MIT) for providing $4.5~\mu$m phase curve data of HD~149026b, to Kevin B. Stevenson (U. Chicago) for phase-resolved spectral data of WASP-43b, and to Daniel Angerhausen (RPI) for updated \emph{Kepler} secondary eclipse analysis. The authors thank Mark Marley for useful suggestions that improved the manuscript. This research has made use of the Exoplanet Orbit Database and the Exoplanet Data Explorer at exoplanets.org.

\footnotesize{
	\bibliographystyle{mn2e}
	\bibliography{JCSreferences}
}

\appendix

\bsp

\label{lastpage}

\end{document}